\documentclass[%
reprint,
amsmath,amssymb,
aps,prl
]{revtex4-2}

\usepackage{graphicx}
\usepackage{dcolumn}
\usepackage{bm}
\usepackage[english]{babel}
\usepackage{hyperref}
\usepackage[mathlines]{lineno}
\usepackage[utf8]{inputenc}
\usepackage{xcolor}
\usepackage{csquotes}
\usepackage{siunitx}

\begin{document}

\title{Higgs modes in  superconducting Titanium nanostructures}

\author{Laurine Marian, Edouard Pinsolle, Samuel Houle, Maxime Durand-Gasselin, Christian Lupien}
\author{Bertrand Reulet}%
\affiliation{Département de Physique and Institut Quantique, Université de Sherbrooke, Sherbrooke, Québec, Canada J1K 2R1}

\date{\today}

\begin{abstract}
We report observations of Higgs modes in superconducting Titanium nanostructures at very low temperature. They appear as anomalies in the microwave complex impedance of the samples revealed by the presence of a dc supercurrent. By varying the sample geometry and contact material, we probe how the Higgs modes are sensitive to the dimensionality of superconductivity, the penetration of the dc and ac current densities in the sample and the dissipation in the contacts.
\end{abstract}

\maketitle

Higgs-Anderson modes (HM) in matter – the oscillation in time of the amplitude of an energy gap – has attracted a lot of attention but suffers from being weakly coupled to the electromagnetic field, thus notoriously difficult to detect\cite{Visibility_HM,Shimano_THz,Light_induced,HMinNIS_AC,HMcoupledtoLCcircuits,HMinJJ,HMJJ2_arxiv,Raman,Raman_recent,Th_Exp_THz_Benfatto,spectroscopHiggs,Higgs_Leggett, Volovik_light_2016}. It has been suggested recently\cite{moor_amplitude_2017} that in a superconductor, the presence of a dc supercurrent $I_s$ could make the mode much easier to detect, by making the coupling to the electric field go from second order\cite{Varma_Review,Shimano_2020,Kubo_dirtylimit} (i.e., proportional to the square of the electric field amplitude $E$) to first order, more precisely, proportional to $I_sE\cos\theta$ with $\theta$ the angle between the electric field and the supercurrent. 
This prediction has been experimentally confirmed in a cm sized thin film of NbN under THz radiation, in the presence of a dc current of several amperes\cite{nakamura_infrared_2019}.

\begin{table*}[htbp]
\scalebox{1.35}{%
\centering
\begin{tabular}{l|c|c|c|c|c|c|c|c|c|c}
\text{Sample} & \text{ Size $L \times w \times t$} & \textbf{$T_{C}$} & \textbf{$R_N$ } & \textbf{$I_C$ } & \textbf{$D \times 10^{4}$} & \textbf{$l$} & \textbf{$2\Delta_0/h$} & \textbf{$\xi$} & \textbf{$\lambda_{L}$} & \textbf{$\lambda_{L}^\perp$}\\ 
 & ($\mu\mathrm{m}\times\mu\mathrm{m}\times \mathrm{nm}$) & (mK) & ($\Omega$) & ($\mu\mathrm{A}$) & ($\mathrm{m}^2 \mathrm{s}^{-1}$) & (nm) & (GHz) & (nm) & (nm) & ($\mu$m)\\
\hline
 WTk & 63 x 10 x 100 & 460 & 7.0 & 450 & 26 & 6.4 & 33.7 & 156 & 514 & 2.64\\
 NTk & 63 x 1.4 x 150 & 495 & 31.5 & 490 & 27.5 & 6.8 & 36.3 & 155 & 485 & 1.57\\
 WTn & 45 x 60 x 15 & 255 & 27 & 76.6 & 5.3 & 1.3 & 18.7 & 95 & 1155 & 89\\
 NTk(Ti) & 48 x 1 x 150 & 520 & 37 & $>$500 & 23 & 5.8 & 38.1 & 140 & 494 & 1.63\\
 NTk(Al) & 63 x 1.8 x 150 & 495 & 28.6 & $>$500 & 23.5 & 5.8 & 36.3 & 144 & 521 & 1.81\\
\end{tabular}
}
\caption{\label{tab: Table1}Properties of the samples. We studied 3 different geometries: wide and thick (WTk), narrow and thick (NTk) and wide and thin (WTn). Samples WTk, NTk and WTn are contacted with a normal metal (Ag), where samples NTk(Ti) and NTk(Al) are contacted with superconductors (respectively Ti and Al). Columns: sample name, dimensions (length $\times$ width $\times$ thickness), critical temperature $T_c$, normal state resistance $R_N$, critical current $I_c$ (at the lowest temperature available), diffusion coefficient $D=(n_Fe^2\rho)^{-1}$ (with $\rho$ the normal state resistivity, $n_{F}=1.35\times10^{47}\mathrm{J}^{-1}\mathrm{m}^{-3}$ the density of states at Fermi level and $e$ the electron charge), mean free path $l=3D/v_F$ (with $v_F$=1.21x$10^{6}\mathrm{m}\mathrm{s}^{-1}$ the Fermi velocity), twice the zero temperature BCS gap $\Delta_0=1.76k_BT_c$ in frequency units, coherence length $\xi=\sqrt{{\hbar D}/{\Delta_0}}$, London penetration depth $\lambda_L=\sqrt{\hbar\rho/\mu_0\pi\Delta_0}$ (with $\mu_0$ the vacuum permeability) and perpendicular London penetration depth $\lambda_L^\perp=\lambda_L^2/t$.\cite{tinkham}}
\end{table*}

In order to pursue the exploration of HM in superconductors, it is highly desirable to generate, control and detect them in circuits, where the electromagnetic wave and dc currents could be controlled at small scale. To do so it is essential to be able to answer fundamental questions such as: how should the dimensions of the structure compare with the relevant length scales of superconductivity, the coherence length and London penetration depth? What are the mechanisms responsible for the decay of the HM, and by what factor are they influenced? Here we provide some answers to these questions by the study of HM in Ti nanostructures at very low temperature and in the microwave domain up to 33 GHz. We have observed the HM in all the samples at all temperatures from well below the critical temperature $T_c$ up to very close to $T_c$. But the visibility of HM varies with geometry and choice of the contacting material.

We have measured five Ti wires of different geometries and with contacts made of different materials, see Table I. We named the three geometries \enquote{WTk}, \enquote{NTk} and \enquote{WTn} which respectively stands for \enquote{wide and thick}, \enquote{narrow and thick} and \enquote{wide and thin}. This allowed us to test different limits of superconductivity: samples WTk and NTk have a thickness of the order of the superconducting coherence length $\xi$ so they can be considered 3D. 
In contrast, the sample WTn has a thickness much smaller than $\xi$ so its electrodynamics may differ from that of a usual 3D BCS superconductor. The critical temperature $T_c$ of all samples is close to the bulk value except for the sample WTn which has much smaller $T_c$, as usually observed in ultrathin films\cite{SC_thin_films}. Another important consideration is the penetration of the electromagnetic field and dc current in the sample, which characteristic length is the London penetration depth $\lambda_L$. Samples WTn and WTk are wider than $\lambda_L$ whereas  sample WTn is narrower. In the former case penetration of the field is restricted to the sides of the sample only.
Samples WTk, NTk and WTn have contacts made of Ag. We named NTk(Al) and NTk(Ti) samples that are geometrically identical to NTk but differ by the material the contacts are made of: while sample NTk is contacted to Ag pads, i.e. a normal metal, NTk(Ti) has Ti contacts and NTk(Al) has Al contact, a superconductor with a gap higher than the one of Ti (the critical temperature of the Al pad is $\sim$\SI{1.1}{K}).

\begin{figure}
\includegraphics[width=7.5cm,height=7.5cm, keepaspectratio]{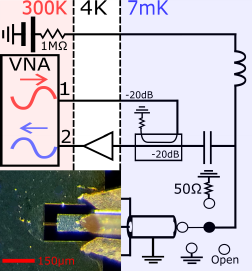}
\caption{\label{fig:schema}Experimental setup. Bottom left: photograph of the picoprobe contacting the Ti wire: the center tip of the probe is connected to the center of the CPW, leading to the Ti wire; the two external tips contact the ground plane.}
\end{figure}

The samples have been fabricated using standard techniques of photolithography, lift-off and metal deposition using e-beam evaporation on oxidized Si substrate. Rectangular Ti thin films are deposited using a first mask of photoresist (see dimensions in Table I). Contacts in Ag or Al of thickness \SI{200}{nm} are then deposited using a second mask. Just before the contact deposition, the surface of the Ti has been etched by ion milling in order to obtain good electrical contact between Ti and the contact layer. For sample NTk(Ti), which is entirely made of Ti, the deposition has been done in one step so there is no interface between contact and sample and the thickness is the same everywhere (\SI{150}{nm}). However, to minimize the contact resistance between the microwave probe (see below) and the Titanium, which oxidizes rapidly when exposed to air, we added an Ag layer on top of Ti contact solely where the contact is made. The contacts have a coplanar waveguide geometry with characteristic impedance $Z_0$=\SI{50}{\ohm}. The transition between the 2D geometry of the sample and 3D geometry of the \SI{50}{\ohm} coaxial cables used in the measurement setup is performed by a microwave probe with \SI{40}{GHz} bandwidth (Picoprobe 40A-GSG-150-T). The position of the probe is adjustable on 3 axis. It is aligned with the sample and the contact is performed under binocular at room temperature. Then the whole sample holder (closed to avoid any influence of parasitic external signals) is installed on the mixing chamber of a dilution refrigerator which can be adjusted in temperature between \SI{7}{mK} and \SI{800}{mK}. 
Au wire bonds between the ground plane of the sample and the copper sample holder insure good thermalization. An optical photography of the sample connected to the probe is presented in Fig.~\ref{fig:schema}.

We have measured the complex impedance $Z(f)$ of all samples placed at ultra-low temperature in the presence of a dc current, where the frequency $f$ varies between \SI{500}{MHz} and \SI{33}{GHz}. 
A schematics of the measurement setup is shown in Fig.~\ref{fig:schema}. Port 1 of a Vector Network Analyzer (VNA) is used as a microwave source that can be swept between \SI{10}{MHz} and \SI{40}{GHz}. This signal is carried down to the sample through lossy coaxial cables, discrete attenuators and a \SI{20}{dB} directional coupler. The wave reflected by the sample is detected by port 2 of the VNA after cryogenic amplification at the \SI{4}{K} stage of the cryostat. The scattering parameter $S_{21}$ (transmission from port 1 to port 2) is measured. Superconducting cables are used between the mixing chamber and the amplifier. The bandwidth of the setup is currently limited by those of the coupler and amplifier. While the bandwidth of the amplifier is nominally 1-\SI{18}{GHz}, it appeared to be usable up to $\sim$\SI{33}{GHz} with lower performance (less gain, more noise). A bias tee is inserted in the circuit in order to inject a dc current in the sample and perform low frequency transport measurements to obtain the critical temperature $T_c$, the normal resistance $R_N$ and the critical current $I_c(T)$, reported in Table I.

The link between the measured parameter $S_{21}$ and the reflection coefficient $\Gamma$ of the sample is complicated because of the strong attenuation along the cables, parasitic reflections on the numerous connectors in the circuit and imperfections of the various components\cite{pozar}. The classical solution to this problem is the use of calibration standards\cite{Calibration_standard}, i.e. samples of well known impedance, the measurement of which allows to correct from all the imperfections of the circuit between the VNA and the sample.  The calibration steps require to replace the sample by three standards, usually an open circuit, ($\Gamma$=+1), a short-circuit ($\Gamma$=-1) and a \SI{50}{\ohm} load ($\Gamma$=0). Since the characteristics of the circuit between the VNA and the sample depend on temperature, it is mandatory to perform the calibration at low temperature. We have implemented such a cryogenic calibration procedure using a microwave switch placed on the base plate of the dilution refrigerator on which the three standards, together with a sample, are connected. Such a procedure allows to bring the calibration plane at the output of the switch, as if the ports of the VNA were placed in the cryostat.

After this procedure we still have to correct the data from the effect of the coaxial cable, the probe and the CPW leading from the switch to the sample. The main effect of this ensemble, rated for \SI{40}{GHz} operation, is to add attenuation $A$ and delay $\tau$. We model the measurement as $\Gamma_1+Ae^{i\omega\tau}\Gamma$, with $\Gamma_1$ the parasitic reflections between the sample and the switch. We measured the attenuation $A$ during a separate cooldown with the probed lifted, i.e. terminated by an open circuit. To correct for the delay $\tau$ we measured the sample in the normal state ($T>T_c$ or $I>I_c$). Even in the normal state the sample cannot be considered a pure resistance. It has a geometric inductance, and the transition from CPW to the sample is accompanied with parasitic inductance and capacitance. We have performed numerical simulations using a finite element software (Ansys HFSS) to calculate these effects. Fitting the measurement against the numerical simulation provides the delay. Then, we take the difference between measurement and fit as the parasitic reflexion $\Gamma_1$. Once the correction is applied, we can extract $\Gamma$ and more importantly the impedance $Z=Z_0(1-\Gamma)/(1+\Gamma)$ of the sample.

\begin{figure}
\includegraphics[width=\columnwidth]{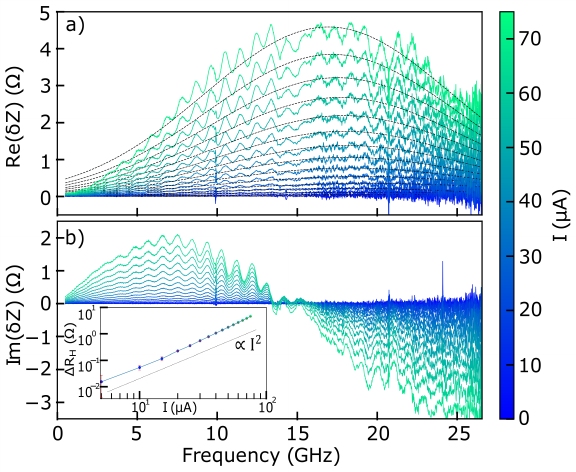}
\caption{\label{fig:dZ} (a): Real and imaginary parts of $\delta Z$  vs. frequency for sample WTn at $T=7$ mK, for various dc currents $I$ between $0$ (blue) and 75 $\mu$A (green), with $I_c=76.6\mu$A. Dotted black lines are Gaussian fits. Inset: peak height of Re($\delta Z$), $\Delta R_{H}$, vs. $I$ in log-log scale. For comparison, the dashed line corresponds to $I^2$.}
\end{figure}

Since the observation of HM relies on the presence of a dc supercurrent, we now focus on the difference $\delta Z(I)$ in the impedance between $Z$ in the presence and in the absence of current: $\delta Z(I)=Z(I)-Z(I=0)$. We show this quantity in Fig.~\ref{fig:dZ} for sample WTn. (a) is the real part and (b) the imaginary part of $\delta Z$. 
As can be seen in Fig.~\ref{fig:dZ}(a), we clearly observe that the presence of supercurrent gives birth to a peak in Re($\delta Z$) vs. frequency. Concurrently a signal centered on the same frequency but anti-symmetric appears on Im($\delta Z$), see Fig.~\ref{fig:dZ}(b). In order to analyze the position, height and width of the peak measured on Re($\delta Z$), we fit the data by a Gaussian centered on frequency $f_H$, of height $\Delta R_H$ and of full width at half maximum FWHM, as in \cite{nakamura_infrared_2019}, see black dotted lines in Fig.~\ref{fig:dZ}(a). We clearly observe that the peak position $f_H\sim$\SI{16.9}{GHz} is independent of $I$. It is close to the expected value $2\Delta_0/h=$\SI{18.7}{GHz}. The height $\Delta R_H$ grows like $I^2$ over more than a decade in current, see inset of Fig.~\ref{fig:dZ}. We did not find any current-dependence in the FWHM within $\sim$5\% at all temperatures. These results agree with those of \cite{nakamura_infrared_2019} and with theory \cite{moor_amplitude_2017}. 

\begin{figure}
\includegraphics[width=\columnwidth,keepaspectratio]{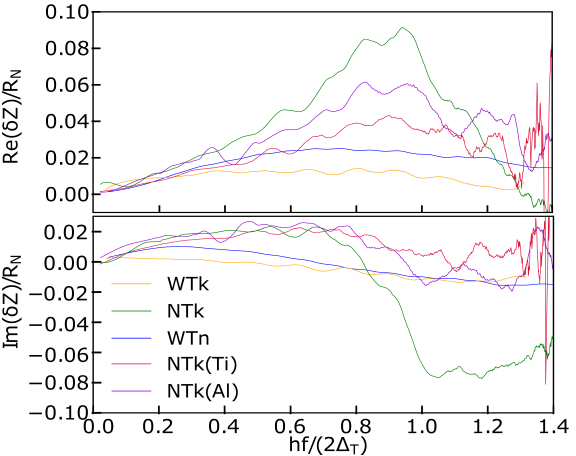}
\caption{\label{fig:Z_Comp} Normalized real  and imaginary parts of $\delta Z$ vs. normalized frequency $hf/(2\Delta(T))$ for all the samples. $\Delta(T)$ is calculated numerically using the BCS integral equation of the superconducting gap at temperature $T$. Green: sample NTk at $T/T_c=0.85$, $I/I_c=0.89$. Violet: sample NTk(Al) at $T/T_c=0.85$, $I/I_c=0.82$. Red: sample NTk(Ti) at $T/T_c=0.85$, $I/I_c=0.83$. Blue: sample WTn at $T/T_c=0.88$ ; $I/I_c=0.85$. Orange: sample WTk at $T/T_c=0.87$, $I/I_c=0.87$.   }
\end{figure}

For samples WTk and NTk, the frequency corresponding to twice the zero temperature superconducting gap was beyond the maximum frequency of our setup. However by increasing the temperature we could lower the gap to bring it in our detection range. We show in Fig.~\ref{fig:Z_Comp} $\mathrm{Re}(\delta Z)/R_N$ as a function of the rescaled frequency for all samples with the same ratio $I/I_c\sim 0.85$: the five samples exhibit the same phenomenon, the apparition of a peak in the presence of a dc supercurrent at a frequency slightly below $2\Delta(T)/h$. This peak is associated to the conversion of the incident electromagnetic wave to HM.
The amplitude of the HM depends on the geometry of the samples: for the wide samples WTk and WTn ($w\gg2\lambda_L$), $\Delta R_{H}/R_N<2\%$, which is similar to what has been reported on macroscopic films \cite{nakamura_infrared_2019}. In contrast, a confining geometry such as NTk ($w\sim2\lambda_L$) strongly enhances the generation of HM where $\Delta R_{H}/R_N$ reaches $\sim 9$\% in similar conditions ($T=0.85T_c$, $I=0.82I_c$). Excitation of the HM requires an overlap of a dc supercurrent density and an ac electric field. This situation is maximized for narrow samples. Note that in our geometry both the dc supercurrent and ac electric field are aligned with the wire ($\theta=0$). The ac field is injected through the contacts and does not penetrate fully the Ti wire if the latter is too wide or thick. This situation notably differs from the experiment on macroscopic film irradiated by a plane wave perpendicular to it \cite{nakamura_infrared_2019}. Samples WTn and WTk are both much thinner than $\lambda_L$. WTn is $6\times$ wider than WTk, yet shows a HM twice larger. This may be due to the 2D character of WTn, or the fact that the penetration of the electromagnetic field is enhanced by WTn being much thinner than WTk\cite{Henkels77} (the relevant length scale for the penetration of a magnetic field perpendicular to the film, $\lambda_L^\perp$, which might be relevant here, is given in Table \ref{tab: Table1}, and indeed the width of WTn is much smaller than that length scale). In any case, the clear observation of HM in sample WTn, see Fig.~\ref{fig:dZ}, shows that being much thinner than $\xi$ is not an issue. 

\begin{figure}
\includegraphics[width=\columnwidth,keepaspectratio]{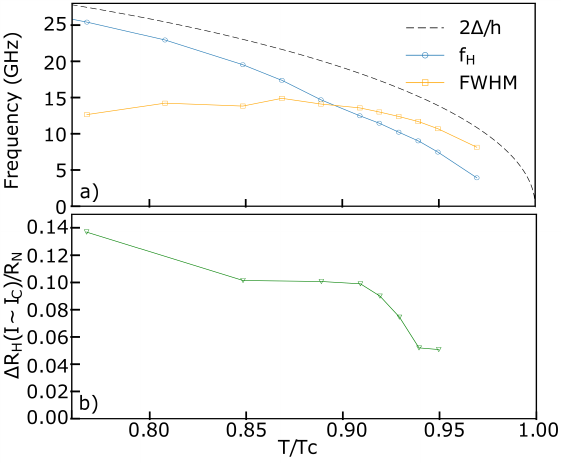}
\caption{\label{fig:Resume} (a) Temperature dependence of the position $f_H$ (blue) and full width at half maximum FWHM (orange) of the HM for sample NTk. For comparison,  $2\Delta(T)/h$ (black dashed line) is the calculated superconducting BCS gap. (b) Temperature dependence of the HM peak height $\Delta R_H(T)$ for the same sample for $I\sim I_c$.}
\end{figure}

We now discuss the temperature dependence of the peak parameters for sample NTk, shown in Fig.~\ref{fig:Resume}. As the peak cannot be well identified and characterized at very low temperature, we focus on temperatures close to $T_c$.
We observe that the peak position $f_H$ (blue line) qualitatively follows the gap $2\Delta(T)/h$ (black), as predicted, but it is always smaller, in particular close to $T_c$. In contrast, the FWHM (orange) seems quite insensitive to the temperature. This result strongly differs from observations on a NbN film\cite{nakamura_infrared_2019}, where the absorption peak has been reported to narrow at low temperature, by a factor $\sim3$ in absolute value between $T_c$ and $0.75\,T_c$. This suggests that in our case the lifetime of the HM is limited by other factors. One possible factor is the presence of the Ag contacts, prone to possible dissipation. In order to explore the possible effect of the material used for the contacts we have made samples of identical geometry with contacts being superconducting with a higher gap (Al) or with the same gap (all the sample made of Ti). The results of Fig.~\ref{fig:Z_Comp} show that Ag and Al contact have identical HM, which rules out the possibility of dissipation in the contacts. With Ti contacts the peak is smaller by a factor $\sim2$ and of comparable width. This might be related to the lack of confinement in Ti sample: the HM generated within the wire, where the dc and ac current densities are high, might leak into the contacts if these are made of the same material as the wire; it cannot escape the wire if the contacts have zero gap or a larger gap.
The temperature dependence of the peak height for $I\sim I_c(T)$ is shown in Fig.~\ref{fig:Resume}b. It is obtained by extrapolating the $I^2$ behavior up to the critical current. We observe that 
$\Delta R_H$ increases at lower temperature, to reach $\Delta R_H(I\sim I_c)/R_N\sim 14$\% for sample NTk at $T=0.75T_c$. It may even increase much more at very low temperature, however $f_H$ becomes too large to extract a reliable value of $\Delta R_H$. For sample WTn, $\Delta R_H(I\sim I_c)/R_N$ strongly increases from 2.6\% at $T=0.85T_c$ up to 17\% at the lowest temperature, see Fig.~\ref{fig:dZ}(a). 

While the position of the HM peak is well understood, and experimentally remarkably robust on all the samples once renormalized by the temperature-dependent gap, the origin of its width, generally attributed to the coupling with quasiparticules is still unclear and would necessitate more investigations.
We have explored the effects of dimensionality, geometry and contact materials on the HM in superconducting Ti nanostructures. By answering questions such as how should be chosen the thickness of the thin film, the width of the wires or the material used for contacts, our study opens the route towards utilizing HM to control and modulate the high frequency properties of superconducting circuits. 
Another avenue would be the direct measurement of HM in the presence of other excitations such as charge density wave or surface plasmon polariton (SSP). Normaly the coupling with plasmon is not possible due to the plasma frequency $\omega_p$ being much higher than 2$\Delta$  but in a recent work \cite{plasmons_higgs} it has been predicted that SPP frequencies in bidimensionnal materials deposited on a superconductor can be lowered enough to couple with the HM.

We thank M. Aprili, C. Bourbonnais, I. Iorsh and D. Sénéchal for fruitful discussions. We are very grateful to G. Laliberté for technical help. This work was supported by the Canada Research Chair program, the NSERC, the Canada First Research Excellence Fund, the FRQNT, and the Canada Foundation for Innovation.

\bibliographystyle{apsrev4-2}
\bibliography{Higgs_Article}
\end{document}